\documentclass[hidelinks]{IEEEtran}
\usepackage{cite}
\usepackage{amsmath,amssymb,amsfonts}
\usepackage{algorithmic}
\usepackage{graphicx}
\usepackage{textcomp}
\usepackage{blindtext}
\usepackage{booktabs}
\usepackage{orcidlink}
\usepackage{color, colortbl}
\definecolor{Gray}{gray}{0.9}
\def\BibTeX{{\rm B\kern-.05em{\sc i\kern-.025em b}\kern-.08em
    T\kern-.1667em\lower.7ex\hbox{E}\kern-.125emX}}
\begin{document}
\title{Subthalamic Nucleus segmentation in high-field Magnetic Resonance data. Is space normalization by template co-registration necessary?}
\author{Tomás Lima \textsuperscript{1,2}\orcidlink{https://orcid.org/0009-0002-1895-9028}, Igor Varga \textsuperscript{2}\orcidlink{https://orcid.org/0000-0002-1085-4312}, Eduard Bakštein\textsuperscript{2}\orcidlink{https://orcid.org/0000-0002-4672-4923}, Daniel Novák \textsuperscript{2}, and Victor Alves \textsuperscript{1}\orcidlink{https://orcid.org/0000-0003-1819-7051}

\vspace{1cm}
\textsuperscript{1}LASI/Algoritmi, University of Minho, Braga, Portugal

\textsuperscript{2}Department of Cybernetics, Czech Technical University in Prague, Prague, Czech Republic

}

\maketitle

\begin{abstract}
Deep Brain Stimulation (DBS) is one of the most successful methods to diminish late-stage Parkinson's Disease (PD) symptoms. It is a delicate surgical procedure which requires detailed pre-surgical patient's study. High-field Magnetic Resonance Imaging (MRI) has proven its improved capacity of capturing the Subthalamic Nucleus (STN) - the main target of DBS in PD - in greater detail than lower field images. Here, we present a comparison between the performance of two different Deep Learning (DL) automatic segmentation architectures, one based in the registration to a brain template and the other performing the segmentation in in the MRI acquisition native space. The study was based on publicly available high-field 7 Tesla (T) brain MRI datasets of T1-weighted and T2-weighted sequences. nnUNet was used on the segmentation step of both architectures, while the data pre and post-processing pipelines diverged. The evaluation metrics showed that the performance of the segmentation directly in the native space yielded better results for the STN segmentation, despite not showing any advantage over the template-based method for the to other analysed structures: the Red Nucleus (RN) and the Substantia Nigra (SN).
\end{abstract}


\section{Introduction}
\label{sec:introduction}
Among other neurodegenerative diseases, Parkinson's Disease (PD) stands out for its big impact on life quality of both patients and caretakers. Its epidemiology is difficult to assess, as diagnosis tends
to be solely based on the physical symptoms in many countries, yet it is estimated that, in Europe, the number lies in 257-1400 PD cases per 100 000 inhabitants \cite{b1}. Nowadays, the most accurate and trustworthy method of diagnose is by Magnetic Resonance Imaging (MRI) of the brain, more specifically by analysing the basal ganglia region \cite{b2}. Besides being an important way of diagnose, MRI also works as a targeting method for Deep Brain Stimulation (DBS), a very effective treatment for advanced PD patients \cite{b3}.

The most common target structure of the DBS in PD patients is the Subthalamic Nucleus (STN), as this basal ganglia structure plays an important role in the pathways of the motor cortex. The traditional targeting strategies of the STN for DBS planning include the manual delineation of the STN borders in T2-weighted (T2w) MRI sequences, the co-registration to brain atlases or the regression based on more easily identifiable neighbour structures such as the Red Nucleus (RN) and Substantia Nigra (SN) \cite{b4}. However, the manual method is very time consuming and requires dedicated time from a specialist for every new segmentation and the other two lack in adaptability, as the STN is a structure that shows anatomical variability across individuals \cite{b5, b6}. Later refinement of the initial segmentation can be done during surgery, using DBS captured microelectrode recordings (MER) \cite{b7}. This procedure can, however, increase the time of surgery and bring discomfort to the patients, so an accurate prior localization of the STN is crucial.

Taking advantage of the power of Deep Learning (DL), most recent automatic segmentation tools show greater adaptability by fitting the learned patterns to each patient's anatomy, while keeping the segmentation times low through the use of the models computational efficiency. Given a dataset of training samples constituted by MRI images and the corresponding segmentations, DL models are able to learn features that allow them to segment future unseen images. Different DL strategies go through the construction of databases of image/segmentation pairs for similarity recognition of new images \cite{b8}, segmentation of various subcortical structures simultaneously to identify useful spatial relations, allowing a more comprehensive analysis of each subject individual anatomy \cite{b9}, multi-atlas registration \cite{b10, b11}, among others. 

Since its introduction in 2015 for a cell segmentation task, UNet has already been successfully applied to the most varied scope of medical and non-medical computer vision problems \cite{b12}. UNet success is related to its capacity of merging low-level features from a classification downsampling path with high-level features of a localization upsampling path \cite{b13}. Several UNet variations can be fond in literature. Possibly the most notably is nnUNet, a framework that allows the simultaneous training of a 2D U-Net, a 3D U-Net and a 3D U-Net Cascade, followed by a process of selection of the best performing out of the three for the case in study \cite{b14}. Brain nuclei segmentation tools have been built using nnUNet powerful adaptability \cite{b15}.

Due to the small dimensions of the STN, it is expected that the automatic segmentation of this structure will benefit greatly from the higher resolution and smaller signal-to-noise ratio that higher-field 7 Tesla (T) MRI can offer \cite{b16}. 

Each medical imaging scan, such as MRI or Computed Tomography (CT), is captured in its own native space. This space is dependent on the scan and its configurations, as well as on the patient positioning. Brain templates are commonly used as a co-registration reference, allowing the comparison across images from different subjects, scans or even imaging modality. One of the most used brain templates in medical research is the Montreal Neurological Institute brain template (MNI).

This study investigates the automated segmentation of STN and the two neighbouring structures RN and SN in high-field 7T brain MRI T1-weighted (T1w) and T2-weighted (T2w) sequences. Publicly available data was used to train and test two different automatic segmentation DL pipelines. The goal was to compare the performance of one architecture relying on co-registration to a brain template with another performing the segmentation directly on native space. Ultimately, this research aims at trailing the path of STN automatic segmentation prior to DBS in high-field MRI, as the technology will become ever more accessible for clinical practice.

\section{Materials and Methods}
\subsection{MRI Data}
Three publicly available 7T brain MRI datasets were chosen for this study. This multi-source selection was done to increase the variability of the data and was based on the availability of both T1w and T2w sequences and on the quality of the images. These datasets are described in the following paragraphs.

The Open Science CBS Neuroimaging Repository (CBS) is a collection of the Max Planck Institute for Human Cognitive and Brain Sciences (MPI-CBS) in Leipzig, Germany, published in 2016. It consists of high resolution 7T T1 and T2 whole-brain scans of 20 healthy subjects of which 12 are female, with an average age of 27.0 ± 3.8 years. The scans were acquired on a Siemens 7T MRI scanner with a 24-channel Nova head coil at 0.5 mm isotropic voxel size. The T1 images were acquired with a MP2RAGE sequence, while T2 images were captured with a multi-echo FLASH sequence. More details about the MRI acquisition can be found in the original paper \cite{b17}.

The 'Atlasing of the basal ganlia' (ATAG) dataset provides whole-brain and reduced field-of-view T1 and T2 scans with high resolution at a sub-millimeter scale. It includes image data from a total of 54 scanned healthy subjects (24 female), organized in three age groups with average age of 23.8 ± 2.3, 52.5 ± 6.6 and 69.6 ± 4.6. The scans were acquired with a 7T Siemens Magnetom MRI scanner using a 24-channel head array Nova coil and consisted of three sequences: a whole-brain T1 with 0.7mm isotropic voxel size, a T1 covering a smaller slab at a finer resolution, and a multi-echo T2w with 0.5mm isotropic voxel size. The T1 images were acquired with a MP2RAGE sequence, while T2 images were captured with a 3D FLASH sequence. Further details about data acquisition of the ATAG data can be found in its publication paper \cite{b18}. Once the STN is only distinguishable in T2 scans, it was decided to discard the ultra-high resolution T1 slab from this study, keeping only the whole-brain T1 and the T2w slab sequences. For the sake of dataset balancing, twenty-five subjects were randomly selected to be part of the study.

The third dataset to be used was a paired dataset (Paired) that contained MRI scans of 10 healthy subjects at both 3T and 7T registered to the same space, which was recently made publicly available by the Biomedical Research Imaging Center (BRIC) of the University of North Carolina. This kind of dataset is an important initiative as it opens the door to studies that bridge concepts between low and high-field MRI images. However, for the sake of this study, only the 7T data was considered. These were acquired with a 7T Siemens Magnetom Terra at a 0.65mm isotropic voxel size, equipped with 32-channel head coils. More detailed acquisition parameters can be found in the original report \cite{b19}. One subject of the Paired dataset was discarded from this study as the images revealed major deformations probably due to the registration process.

\subsection{Training and Test Subsets}
A subset composed of subjects from the three selected datasets was created for testing. The subjects of this subset were not used for training nor validation, thus ensuring that the performance assessment of the models would only be done on previously unseen data. To create the test subset, a 3 to 1 ratio between number of training and test subjects for each dataset was sought. Therefore, five, six and two subjects were randomly selected from the CBS, the ATAG and the Paired datasets, respectively. The distribution of the number of subjects per training/test subsets and per dataset is shown on Table \ref{tab1}

\begin{table}
\centering
\caption{Number of subjects per dataset and division in training and testing subsets.}
\setlength{\tabcolsep}{3pt}
\begin{tabular}{|c|c|c|c|}
\hline
Dataset& 
Train subset& 
Test subset&
Total\\
\hline
CBS& 15& 5&20\\
ATAG& 19& 6&25\\
Paired& 7& 2&9\\
\hline
Total& 41& 13&54\\
\hline
\end{tabular}
\label{tab1}
\end{table}

\subsection{Labels Generation}
\begin{figure}[!t]
\centerline{\includegraphics[width=0.6\columnwidth]{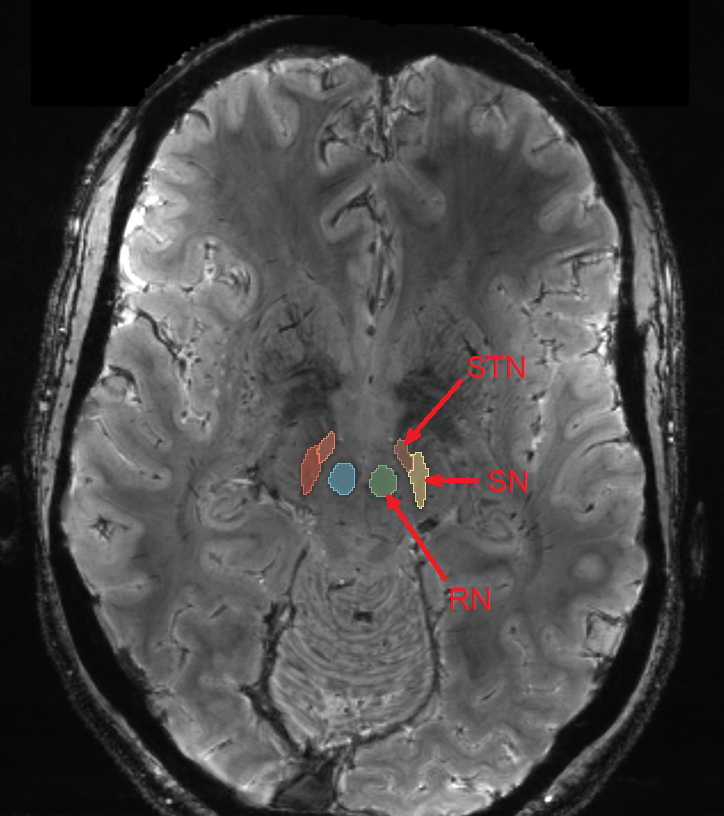}}
\caption{Example of a segmentation map output by the pBrain pipeline overlayed on top of the corresponding T2w image.}
\label{pbrain_example}
\end{figure}

None of the used datasets had ground-truth labels of the target structures. Therefore, labels were generated using pBrain tool \cite{b20} for the CBS and the Paired datasets. pBrain pipeline outputs a segmentation mask with six labels - the three nuclei (STN, SN and RN) in each hemisphere. An example of the output given by the pBrain tool is shown in Figure \ref{pbrain_example}, where the output segmentation map with the six labels has been overlayed on top of the input T2w image. As this tool needs a whole-brain T2 image in order to generate the segmentations, it was not possible to create labels for the ATAG dataset, as this only contained a slab of the T2 sequences. For that reason, all the twenty-five subjects of the ATAG dataset were manually labeled by a trained specialist using the ITK-Snap software \cite{b21}. All subjects of the test subset were manually labeled using ITK-Snap, as these would serve as the ground-truth for the automatic segmentation performance evaluation. In order to match the labels of pBrain, the same six segmentation labels were generated manually. The segmentation maps of each subject were validated by two other specialists, both the pBrain and the manually generated. All labels were generated in the native space of each subject T2w image.

\subsection{Template-Based Automatic Segmentation - Method I}
The first automatic segmentation method relied on the co-registration of the MRI images to the high-resolution MNI Nonlinear Asymmetric version of 2009 (MNI152 NLIN 2009b) brain template \cite{b22}. The main advantage of this strategy is the normalization of the model input images to a common space, which facilitates the construction of a common Region Of Interest (ROI) and reduces the risk of major localization errors.

\begin{figure}[!t]
\centerline{\includegraphics[width=\columnwidth]{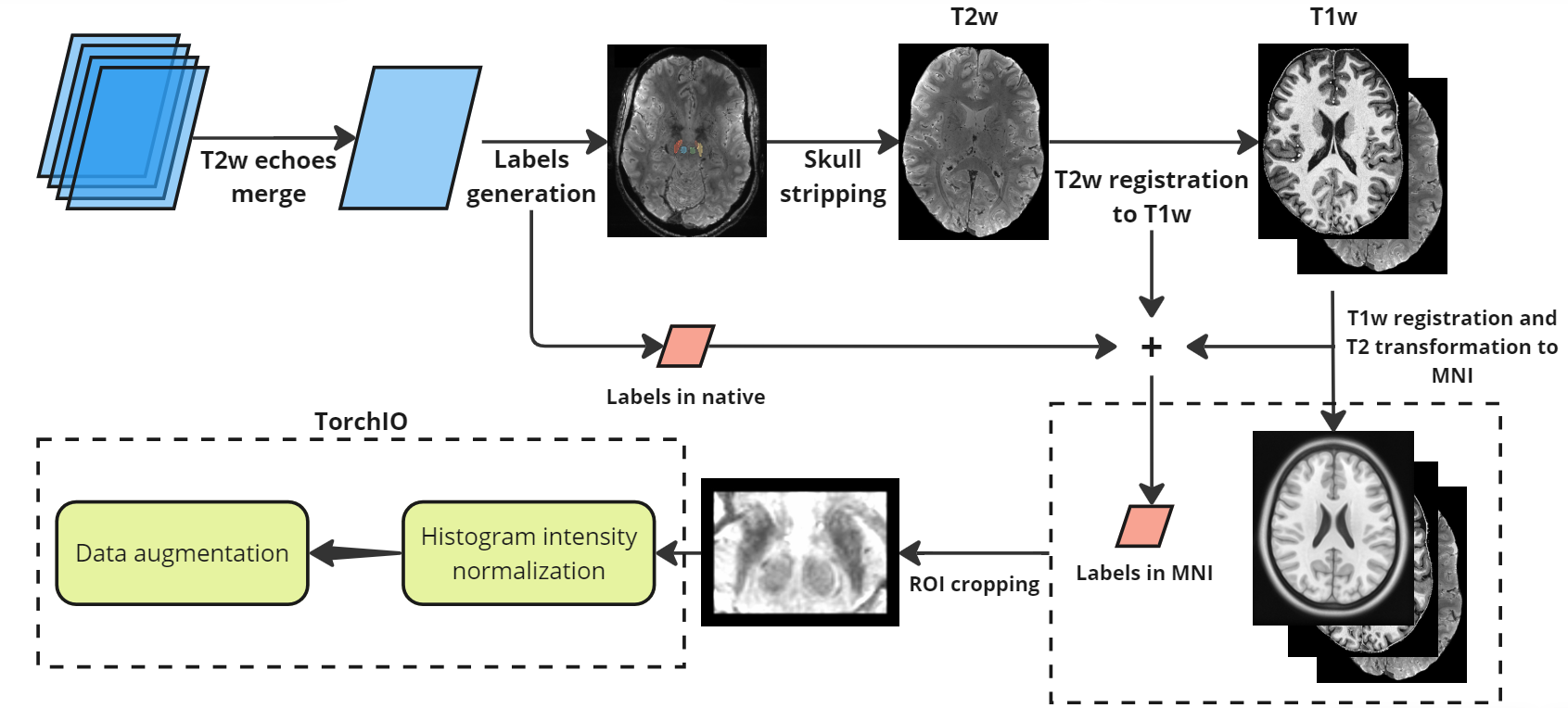}}
\caption{Overall pre-processing pipeline of the MNI based automatic segmentation method.}
\label{Method_I_preprocess}
\end{figure}

The full pre-processing pipeline of Method I is illustrated in Figure \ref{Method_I_preprocess}, which comprises all the steps from the raw dataset images until the final stage with the data ready for model input. These contain the following steps:
\begin{enumerate}
    \item Merging of the multi-echo FLASH sequences, by averaging the intensity values across the different echoes.
    \item Skull stripping using SynthStrip tool \cite{b23}.
    \item Affine co-registration of T2w images to T1w, followed by affine co-registration of the T1w images to the MNI template. The skull-stripped T1 version of the MNI152 NLIN 2009b was chosen as T1 images show higher contrast definitions between structures, which allows the registration models to achieve a better overlay between images. The Elastix toolbox \cite{b24} was use for every registration task. This package allowed the application of the T1w to MNI transformation to the T2w image registered in T1w, so that both T1w and T2w could be in MNI space. This is also why affine registration was preferred over non-linear registration. Elastix also allowed the computation of the inverse transformation of every registration, so that all images and segmentation outputs could be brought back to native space. Using Elastix, all this could be automatised by writing simple script blocks.
    \item Application of T2w to T1w and T1w to MNI transformations to the generated labels.
    \item Labels based ROI cropping - with all subjects registered in the MNI space, the maximum and minimum values for each 3D axis coordinates across all registered training labels were stored. A ten voxels margin was added to these values in order to construct a cubic ROI mask that surrounds the target structures in the MNI space.
    \item Histogram intensity normalization and data augmentation (rotation, translation along the sagittal plane, scaling and noising) using TorchIO framework \cite{b25}. Since the model will work with MNI registered images, data augmentation was restricted to 2º of rotation, 2mm of translation and 0.05 factor of scaling, representing possible imperfections of affine registration to MNI space.
\end{enumerate}

The pre-processed images and corresponding labels of the training subset were used to train a 3D full-resolution nnUNet with a 5-fold cross validation and using nnUNet's most recent residual encoder presets \cite{b26}.

A post-processing pipeline was created in order to bring the nnUNet predictions back to native space. The inverse transformations of the pre-processing registrations were applied to the labels following the method described by Metz et al. \cite{b27}, and the resulting labels in native space were Gaussian smoothed. The smoothing was necessary because of the jagged boundaries that arose from the need to set the interpolation order to 0 when transforms are being applied to label maps. 

\subsection{Native Space Automatic Segmentation - Method II}
The second automatic segmentation method performed the segmentation directly in the native space, avoiding the need to perform image registrations. The main advantages are then the faster computation (no registrations and no smoothing), the possibility to perform more aggressive data augmentation since it is not restricted by a brain template, and the non-dependency on third-party medical imaging registration tools that might introduce misalignment errors on themselves. Since there is no space normalization step through registration to a template, this method required a localization step (Localizer) before performing the segmentation. This was a simple regression model responsible for locating the combined center of mass of the target structures (CoM). The regression network had two input channels (T1w and T2w images as input) and the output layer had three channels, one for each spatial dimension. With a shape standardized ROI constructed around each subject’s individual CoM, the second Segmenter network (nnUNet) segments the target structures in a similar manner to the one of the Method I. 

\begin{figure}[!t]
\centerline{\includegraphics[width=\columnwidth]{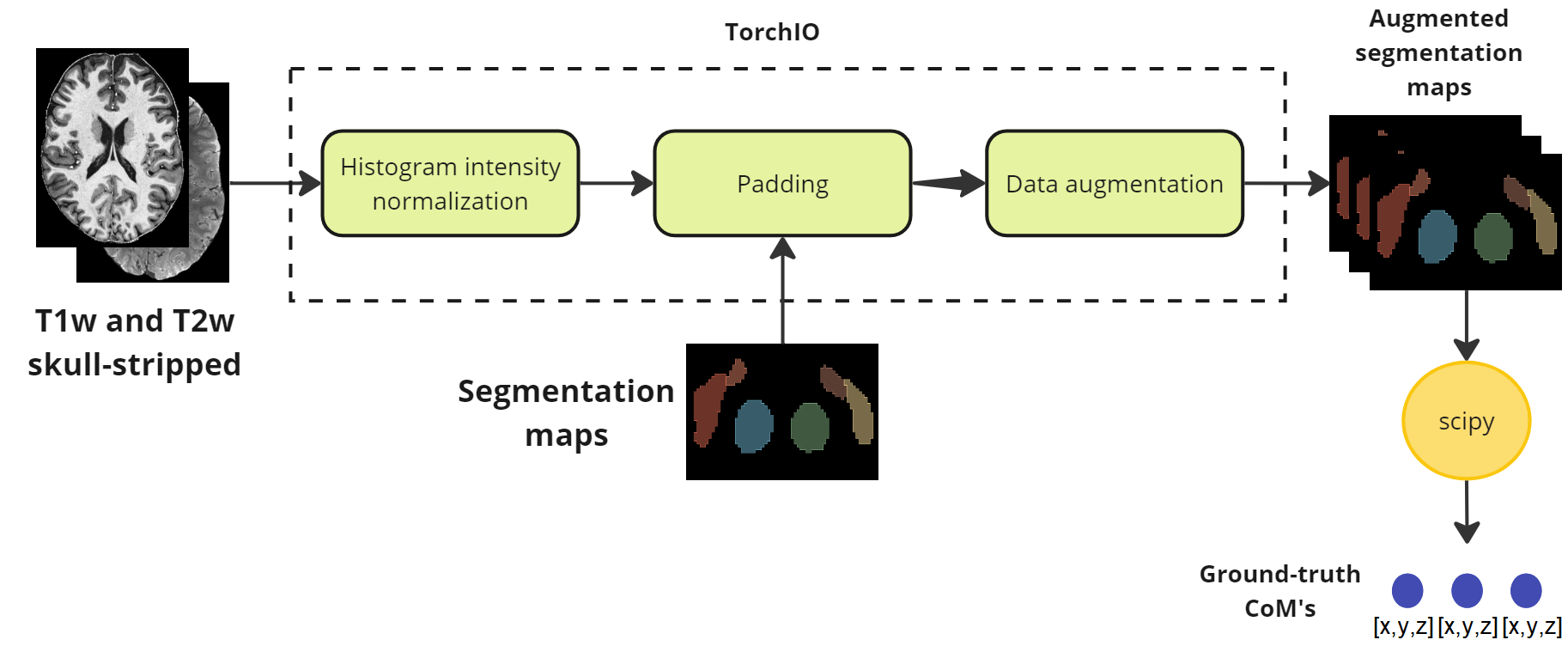}}
\caption {Overall pre-processing pipeline of Method II architecture.}
\label{localizer_preprocessing}
\end{figure}

\begin{figure}[!t]
\centerline{\includegraphics[width=\columnwidth]{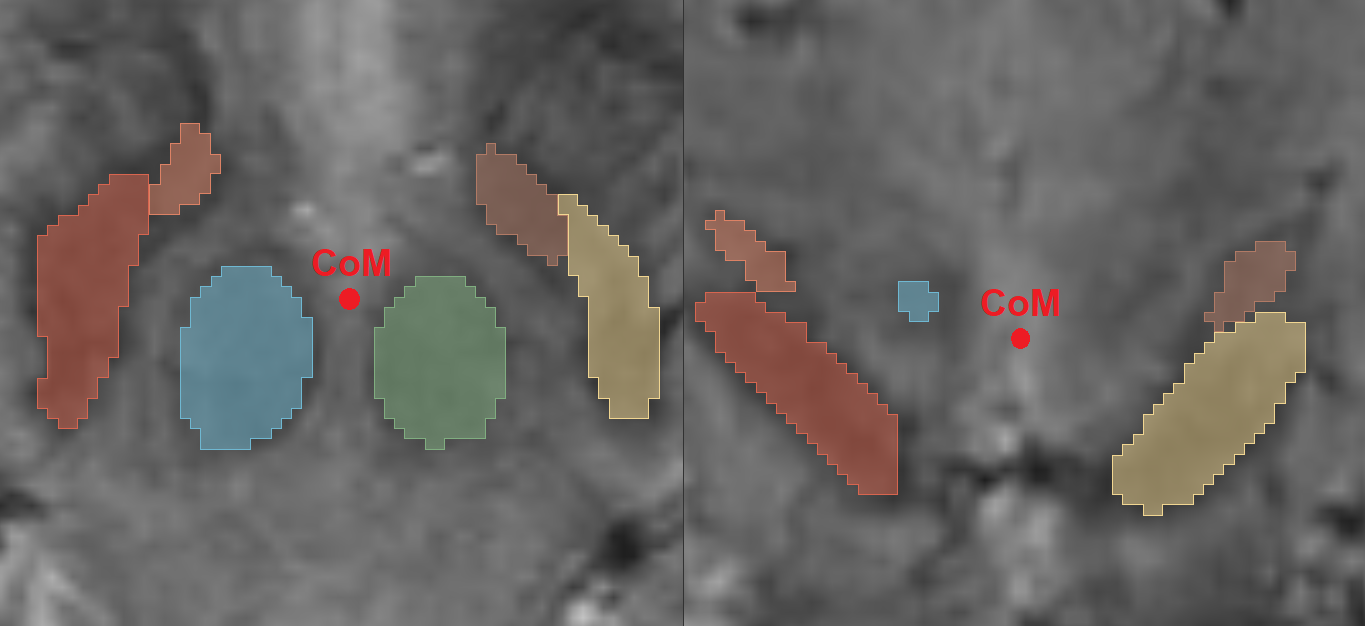}}
\caption {Example of what the CoM of a segmentation map represents in views of the axial and coronal planes on a T2w image with overlay of a segmentation map with the six different labels.}
\label{com_example}
\end{figure}

The two first pre-processing steps were similar to the ones of Method I. Figure \ref{localizer_preprocessing} depicts the remaining Method II specific steps, being the full pre-processing pipeline as follows:

\begin{enumerate}
    \item Merging of the multi-echo FLASH sequences, by averaging the intensity values across the different echoes.
    \item Skull stripping using SynthStrip tool \cite{b23}.
    \item Histogram intensity normalization - performed with TorchIO's \emph{Histogram Standardization} after assembly of the dataset.

    \item Padding of T1w, T2w and labels images - since images were from different datasets, it was necessary to ensure that all of them had the same shape before input into any model. Padding was set to match the biggest value in each dimension across the three datasets.

    \item Data augmentation - the same techniques as in Method I were applied, but with bigger degrees of freedom (15º degrees for rotation, 15mm for translation and scaling factors up to 0.2). 

    \item Computing of CoM for each instance of the augmented training data and the test subset - the combined Center of Mass of each subject's segmentation map was computed using the function \emph{center of mass()} of the \emph{scipy.ndimage} multidimensional image processing package. An example of what the CoM represents is illustrated in Figure \ref{com_example}. Before the CoM computation, all different labels of each subject's segmentation map were converted to 1 (binary image), so that the different values of each label would not bias the CoM position.
\end{enumerate}

Using the pre-processed images and the CoM's, a regression model was trained to perform the prediction of the CoM. The model was trained using the 41 subjects of the train subset and tested in the 13 subjects of the test subset. The error along each axis were averaged in order to determine the size of the ROI to be created around the CoM. According to the results obtained, which are listed in detail in the Results section, we opted for a ROI with dimensions of [78, 72, 60] in order to accommodate the entirety of the target structures and to account for the Localizer errors.

\begin{figure}[!t]
\centerline{\includegraphics[width=\columnwidth]{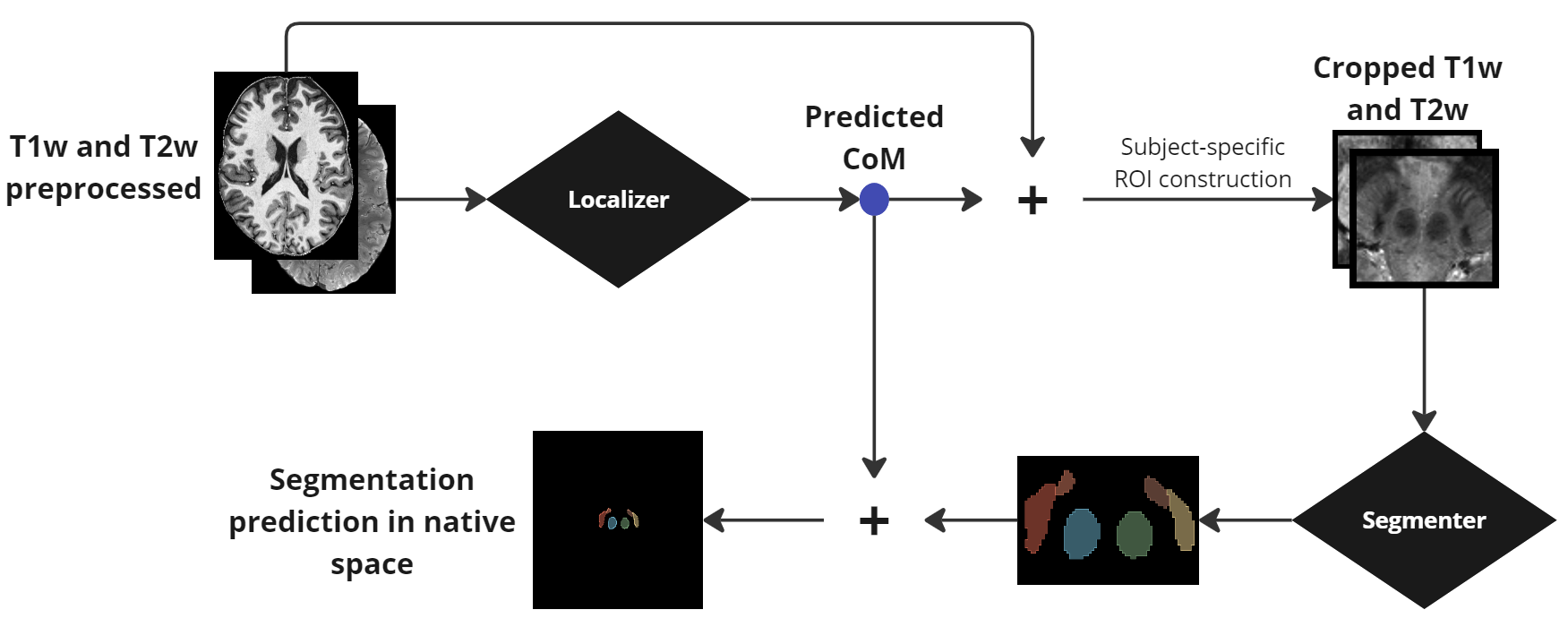}}
\caption { Localizer+Segmenter architecture workflow diagram after data pre-processing.}
\label{localizer_segmenter_diagram}
\end{figure}

Similarly to Method I, a nnUNet was trained to produce the segmentation, maintaining the same parameters. The output segmentation maps of nnUnet were in the same space as the ROI images. Therefore, it was necessary to bring them back to the original image space by centering them at each subject's stored CoM coordinates followed by cropping to match the corresponding shape of the T2w original images of each dataset. The diagram of Figure \ref{localizer_segmenter_diagram} illustrates the previously described steps inserted on the global workflow of the Localizer+Segmenter (Method II) architecture after data pre-processing.

\subsection{Evaluation of the Automatic Segmentations}
The previously described evaluation metrics were apllied in the native space, once every DBS procedure is planned in the patient's native space. In order to evaluate the performance of both methods, Dice Similarity Coefficient (Dice Score) and two surface area and volume metrics were computed. The Dice Score quantifies the overlap between two sets of binary segmentation results, providing an intuitive and interpretable measure of similarity. It is defined as:

\begin{equation}
    \text{Dice Score(A,B)} = \frac{2 \cdot |A \cap B|}{|A| + |B|}
\end{equation}

\noindent
where \(A\) and \(B\) are the sets of voxels belonging to the predicted and ground truth segmentations, respectively. The Dice score ranges from 0 to 1, where higher scores indicate better segmentation performance, with values closer to 1 suggesting a high degree of similarity between the predicted and ground truth labels.

The volume and surface area metrics were defined as:
\begin{equation}
\text{Volume Metric(m,p)}= \frac{1}{N} \sum_{i=1}^{N} \left( \frac{|V_{\text{p},i} - V_{\text{m},i}|}{V_{\text{m},i}} \times 100 \right)
\end{equation}

\begin{equation}
\text{Surface Area Metric(m,p)} = \frac{1}{N} \sum_{i=1}^{N} \left( \frac{|S_{\text{p},i} - S_{\text{m},i}|}{S_{\text{m},i}} \times 100 \right)
\end{equation}

\noindent
where \( V_{\text{p},i} \) and \( V_{\text{m},i} \) denote the predicted and manual volume values for the \(i\)-th sample of the test subset of N subjects, respectively, and \( S_{\text{p},i} \) and \( S_{\text{m},i} \) denote the predicted and manual surface area values for the \(i\)-th sample of the test subset of N subjects, respectively. The relative value of the differences was chosen to remove the potential bias that could exist due to anatomical variability of the structures dimensions across subjects.

\section{Results}
The constructed methods were applied to the test subset, constituted by images of subjects that had not been seen during training of the models, neither for training nor validation.

\begin{figure}[!t]
\centerline{\includegraphics[width=0.8\columnwidth]{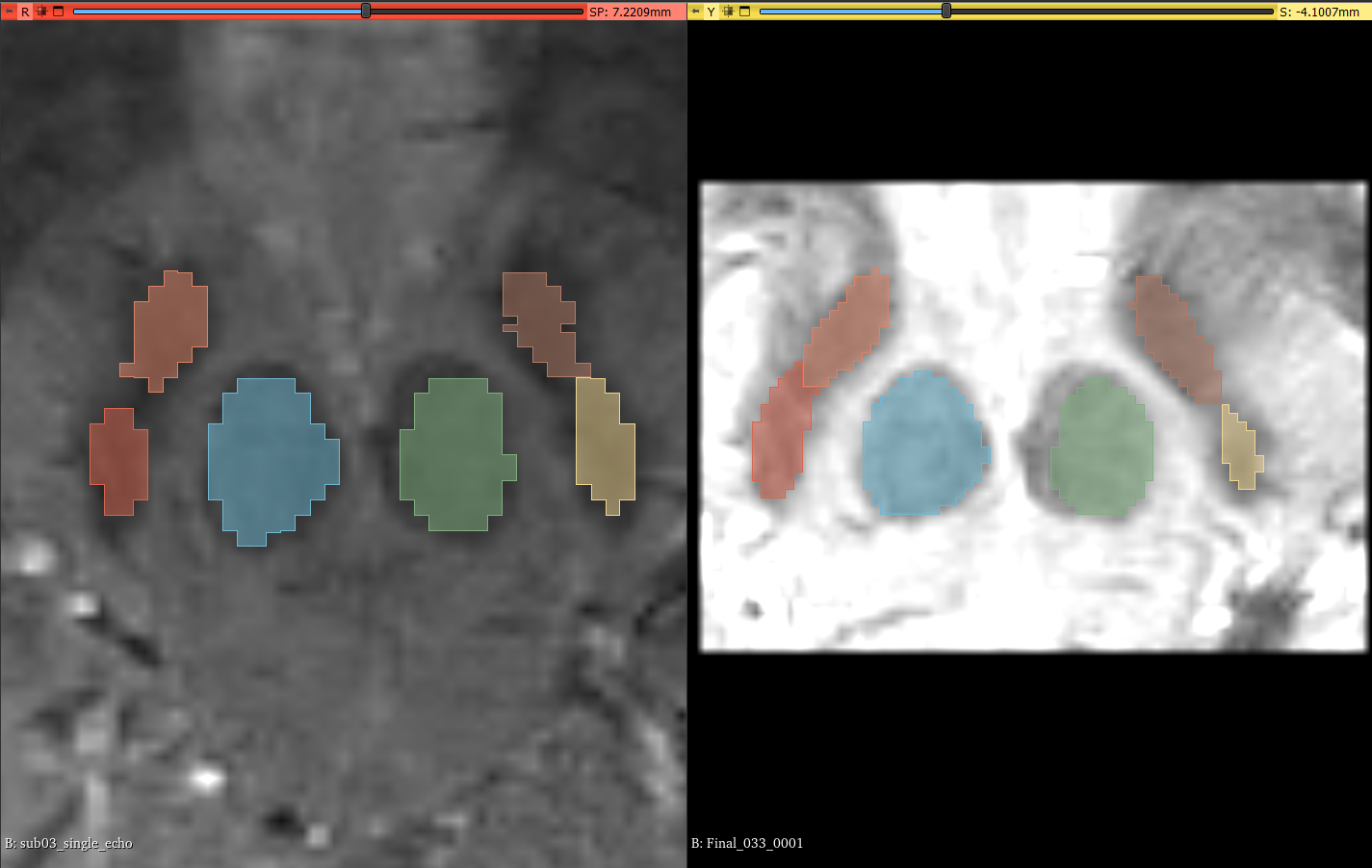}}
\caption {Comparison between the raw output of nnUNet segmentation in the MNI space registered and pre-processed T2w image (right) and the registered to native space and Gaussian smoothed same output (left).}
\label{method_I_labels_before_and_after}
\end{figure}

The average Dices of each of the six structures from both methods are detailed in Table \ref{table:Dices_scores}. As it can be seen from the results obtained for the overall test subset, both methods had a very similar performance for all the structures, with exception of the STN where Method II performed considerably better in both hemispheres.  Method I achieves significantly lower Dice scores for the left STN (STN\_l) than for the right STN (STN\_r). This fact may be related to the need of inverse transforming the labels to native space during the post-processing step, which causes deformation and a jagged effect to the usual smooth labels of nnUNet. This effect is clearly demonstrated in Figure \ref{method_I_labels_before_and_after}, where the rough border effect is visible in the native space registered labels, even after Gaussian smoothing post-processing. This effect will still be analysed in more detail by further metrics.

\begin{table}
\centering
\caption{Dice scores between the predictions of both methods and the manual labels for each of the segmented structures in the test subset.}
\begin{tabular}{c|cc}
\toprule 
Structure & \multicolumn{2}{c}{Dice Score} \\
\midrule
{}   & Method I & Method II  \\ \hline
RN\_r & \textbf{0.877} & 0.868 \\ \hline
SN\_r & \textbf{0.817} & 0.810 \\ \hline
\rowcolor{Gray}
STN\_r & 0.693 & \textbf{0.714} \\ \hline
RN\_l & \textbf{0.904} & 0.894 \\ \hline
SN\_l & 0.804 & \textbf{0.811} \\ \hline
\rowcolor{Gray}
STN\_l & 0.631 & \textbf{0.718} \\ \hline
\bottomrule
\end{tabular}
\label{table:Dices_scores}
\end{table}

The values obtained for the volume and surface area based evaluation metrics given by Equations (2) and (3), respectively, are presented in Table \ref{table:Volume_metrics} and Table \ref{table:area_metrics}. The analysis of these results shows once more the fairly similar performance of both models for the RN and SN, while for the STN Method II shows a better performance. It is worth of notice that Method I performed consistently better than Method II on the volume metric, even by almost seven percentage points in some of the non-STN structures. This over-performance is, however, not verified on the area metric, where Method II even performs better in three out of the four non-STN structures. This finding is once more related to the jagged effect of Method I predicted labels due to the back-to-native registration process (Figure \ref{method_I_labels_before_and_after}).

\begin{table}
\centering
\caption{Average absolute difference between the predicted and manual segmentation volumes in percentage of the manual segmentation volumes. Volume metric given by Equation (2).}
\begin{tabular}{c|cc}
\toprule 
Structure & \multicolumn{2}{c}{Volume Metric} \\
\midrule
{}   & Method I & Method II  \\ \hline
RN\_r & \textbf{13.93\%} & 15.06\%  \\ \hline
SN\_r & \textbf{15.91\%} & 20.08\%  \\ \hline
\rowcolor{Gray}
STN\_r & 28.44\% & \textbf{22.87\%} \\ \hline
RN\_l & \textbf{13.42\%} & 15.03\% \\ \hline
SN\_l & \textbf{13.72\%} & 20.24\% \\ \hline
\rowcolor{Gray}
STN\_l & 31.18\% & \textbf{19.69\%}\\ \hline
\bottomrule
\end{tabular}
\label{table:Volume_metrics}
\end{table}

\begin{table}
\centering
\caption{Average of the absolute values of the surface area metric between the predicted segmentations of each method and the manual labels, given by Equation (3).}
\begin{tabular}{c|cc}
\toprule 
Structure & \multicolumn{2}{c}{Surface Area Metric}\\
\midrule
{}   & Method I & Method II   \\ \hline
RN\_r & 12.87\% & \textbf{10.15\%} \\ \hline
SN\_r & 10.72\% & \textbf{9.70\%} \\ \hline
\rowcolor{Gray}
STN\_r & 27.73\% & \textbf{16.76\%} \\ \hline
RN\_l & \textbf{10.98\%} & 12.41\% \\ \hline
SN\_l & 16.01\% & \textbf{13.26\%} \\ \hline
\rowcolor{Gray}
STN\_l & 34.11\% & \textbf{15.89\%} \\ \hline
\bottomrule
\end{tabular}
\label{table:area_metrics}
\end{table}

\begin{figure}
  \centering
  \begin{tabular}{@{}c@{}}
    \includegraphics[width=.8\columnwidth]{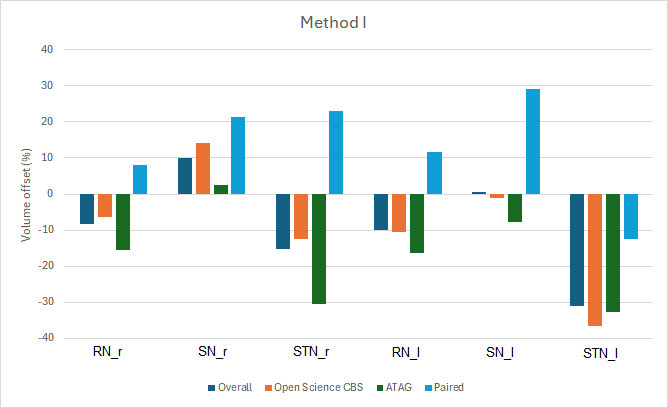} \\[\abovecaptionskip]
  \end{tabular}
    (a)
  \vspace{\floatsep}

  \begin{tabular}{@{}c@{}}
    \includegraphics[width=.8\columnwidth]{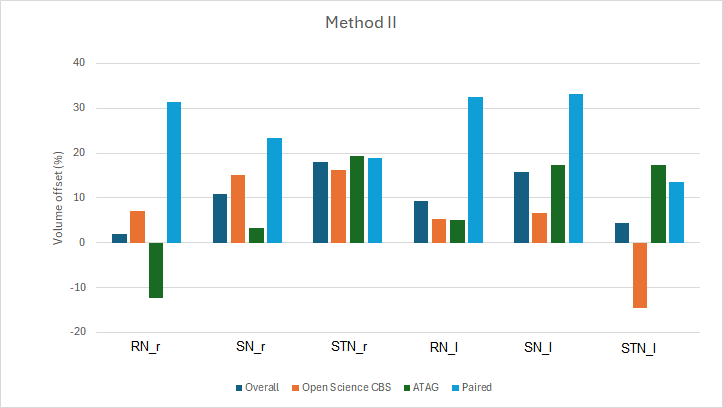} \\[\abovecaptionskip]
  \end{tabular}
    (b)
  \caption{Average volume offsets of each segmented structure by Method I (a) and Method II (b) on the overall test subset and of each individual dataset. The volume differences were computed by subtracting the volume of the manual label to the volume of the predicted label. Positive offsets correspond to over-segmentation and negative values to under-segmentation.}
  \label{volume_offsets}
\end{figure}

The computation of the volume of the segmented structures allowed for the analysis of potential bias of the models to segment in excess or defect the different structures. The histograms of Figure~\ref{volume_offsets} represent the average volume differences of each segmented structure between the automatic labels and the manual label, in percentage of the manual label volume, similarly to Equations (2) and (3), but by averaging the raw difference instead of the absolute value. This allows to detect if the segmentation error is done by over or under-segmenting. The analysis is again done for the overall test subset and individually for each dataset. Figure~\ref{volume_offsets}(a) represents the results of Method I, whereas the results of Method II are shown in Figure~\ref{volume_offsets}(b). The offsets were computed by subtracting the volume of the manual labels to the volume of the predicted, meaning that positive values represent a prediction larger than the real, and negative values stand for automatic smaller predictions than the real. 

An overview of Method I's results shows that this model achieved, on average, volumes closer to the ground-truth. However, the predicted STN labels' volumes were significantly smaller than the real, being the difference between the volumes more than 30\% of the real volume. It is also worth of notice the tendency of the model to segment in excess all the structures in the Paired dataset. This again is related to the lower resolution of its images. Lower resolution means that the target structures appear in less slices. However, the model was trained predominantly with higher resolution images, meaning that it expects the structures to appear in a bigger number of slices. This causes the over-segmentation of the structures. As for Method II, the norm was a segmentation by excess of all the structures, but the difference between the predicted and real STN volumes never exceeded 20\% of the real STN volume. Once again, the Paired dataset was segmented in excess (in this case with even higher discrepancies than the other datasets), for the same reasons as Method I.

\begin{figure}
  \centering
  \begin{tabular}{@{}c@{}}
    \includegraphics[width=.7\columnwidth]{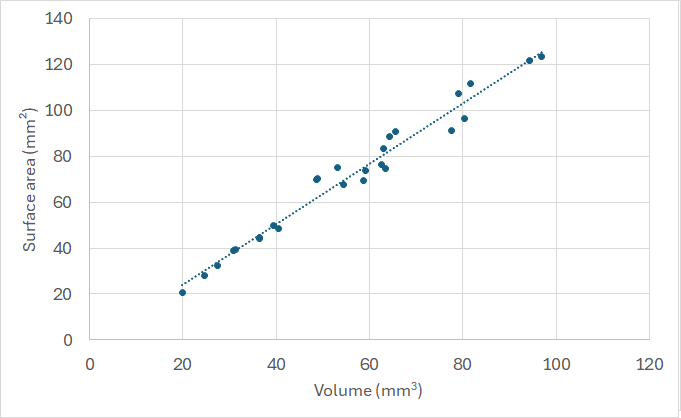} \\[\abovecaptionskip]
  \end{tabular}
    (a)
  \vspace{\floatsep}

  \begin{tabular}{@{}c@{}}
    \includegraphics[width=.7\columnwidth]{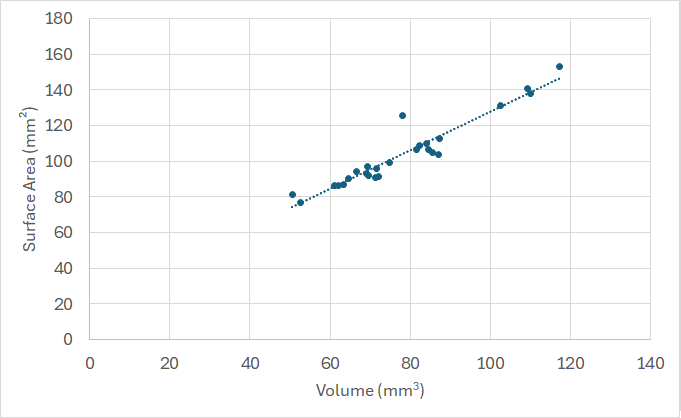} \\[\abovecaptionskip]
  \end{tabular}
    (b)
    \vspace{\floatsep}

    \begin{tabular}{@{}c@{}}
    \includegraphics[width=.7\columnwidth]{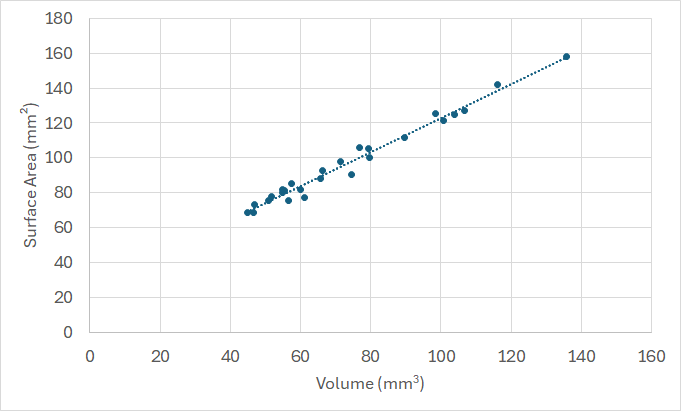} \\[\abovecaptionskip]
  \end{tabular}
    (c)
  \caption{Volume (V) and surface area (A) ratio of left and right STN segmentations of the test subset. (a) Predicted labels of Method I, with linear approximation equation given by $y = 1.3197x - 2.2947$. (b) Predicted labels of Method II, with linear approximation equation given by $y = 1.0858x + 19.401$. (c) Manual labels, with linear approximation equation given by $y = 0.9776x + 25.142$.}
  \label{VA_segmentations}
\end{figure}

As the main target of this study is the STN, the volume to surface area ratios of the STN manual labels and of both methods STN predicted labels were computed and are displayed in Figure~\ref{VA_segmentations} scatter plot graphs. The linear regression approximation to the values is also shown and its equation is detailed in the description. In practice, the volume increases assymptotically with the cube of the length, while surface area increases assymptotically with the square of length, i.e. the proper relation between volume and surface area would follow $A = V^{\frac{2}{3}}$, where A represents the surface area and V represents the volume. However, for the dimensions of the studied STN, the proportion resides still in the linear component of this relation, thus being linear regression a good fit for the sake of simplification. As it is visible, and confirmed by the incline of the linear equations, Method II achieved volume to surface area ratios closer to the ground-truth. The incline of the linear regressions was of 0.9776 (mm\textsuperscript{2}/mm\textsuperscript{3}) and 1.0858 for the manual and Method II STN labels, respectively, representing a difference of 0.1082, while the difference between the inclines of the manual and Method II labels was of 0.2339, meaning that for the same volume, the surface area was over 23\% bigger than it should. This result is the best proof of the jagged effect that the inverse transformations back to native space have in the predicted labels, as it is the cause of the increase of the surface area without greatly affecting the total volume of the segmentation.

\section{Discussion}
Two different methods were described for the automatic segmentation of the STN in high field MRI scans. This task poses itself as one very important step for the planning of DBS, a PD treatment with high rates of success.

The constructed brain template-based segmentation architecture had a worse performance than the non MNI-based model in terms of accuracy for the main target structure of this study, the STN. Possibly due to the lack of variability imposed by the registration to a brain template, it seemed to have more difficulties identifying the boundary between the STN and SN and on generalizing to new data. However, this problem might be solvable with access to more data for more robust training of the model. Moreover, most of the accuracy was lost in the inverse transformation from MNI back to the native space (labels jagged boundaries effect), where the segmentation has to be evaluated. This method did, nevertheless, perform very stably for the other studied basal ganglia nuclei. Therefore, it is expectable that with some refinement of the pipeline registration nuances also the segmentation of the STN can achieve accurate results.

Method II, which performed all its steps in the native space, achieved better segmentation of the harder to segment STN structure. This is mainly due to the possibility it opens to perform more aggressive data augmentation which allows the model to better learn how to identify the low contrast STN borders. Also, it did not require the space transformation of labels, a procedure that affects their quality. This native space based method generalized better to new unseen high-field 7T MRI data.

The use of datasets from different sources containing MRI images acquired by distinct scanner and at different resolutions highlighted the importance of the pre-processing steps when the goal is to create a global automatic segmentation solutions. The multi-resolutions challenge might be addressed with up-sampling (or down-sampling) steps prior to model input, or with the use of bigger and more balanced datasets for training of the models in order to achieve more robustness.

\section{Conclusions}

This study showed evidence that the step of co-registration of MRI images to a template is not entirely necessary to achieve state-of-the-art automatic segmentation performance of the STN in high field brain MRI. Moreover, the construction of algorithms that perform the segmentation in the images native space may allow more aggressive data augmentation techniques, which can be an advantage in cases where less data is available.

Future studies may seek the adaptability of such methods to lower field MRI data (still the most broadly used in clinical practice), on an attempt to prove if the sharper patterns learned in 7T images can be useful at lower fields. Such enterprise might require the adaptation of the training data, either by integrating 1.5T or 3T images, or by simulating such images from 7T by downsampling or intensity adjustment steps.

\end{document}